# THE MOST CONSIDERED TYPE OF STUDENT CHARACTERISTICS BY PRIMARY SCHOOL TEACHERS

Dwiwarna[1] and Raditya Bayu Rahadian[2]

[1]Education and Cultural Offices of South Bangka, Islands of Bangka Belitung, Indonesia
[2]Department of Instructional Technology, Graduate School, Yogyakarta State University, Yogyakarta, Indonesia

## ABSTRACT

*This study aims to determine what type of student characteristic most considered by teachers in primary school and their influence on student learning achievement. For this purpose, 37 class teachers of grade 4th to 6th from seven primary schools were the respondents. Categorizing the level of identification students characteristics by the teacher using a questionnaire on the application of student characteristics that is deepened by the study of learning planning documents intended to provide an overview of the types of students characteristics that are considered most important to be done by primary school teachers. The results show that the level of student intelligence is the type of student characteristics most considered by primary school teachers. By comparing the value of student achievement it can be seen that the learning developed by paying attention to students intelligence levels are better than learning that does not pay attention to it.*

## KEYWORDS

*Student Characteristics, Learning Achievement, Primary Schools*

## 1. INTRODUCTION

Student characteristic is one of the conditions of the instructional variable which is occupied an important position that leads to learning achievement. The kind of student characteristic is intelligence levels, prior knowledge, cognitive styles, learning styles, motivation, and sosio-culture [1]–[3]. On the instructional structure, it was connecting with the instructional management strategy. Management strategy is elemental methods for making decisions about which organizational and delivery strategy components to use when during the instructional process. They include such conciderations as for how to individualize the instruction and when to schedule the instructional resources [3], such us; selecting the instructional component, organizing delivery of content, development instructional strategy and media, delivering matters and motivational management, and managing instructional activities [4]. All of this must be considered by the teachers before they make instructional planning that suitable with the student characteristics for the equitable and meaningful learning.

Due to many types of student characteristics, it often comes up with the questions of teachers, especially in primary schools, such as: which one is more important? Should everything be identified? Can I choose one or more that is more urgent to pay attention to? Which is most influential in improving learning outcomes?

Ideally, all types of student characteristics must be considered by the teachers because the variety of student characteristics influence the selection of other instructional method variables [5]. However various limitations causing them not to study in its entirety and depth. With the result that, need an effort to give a guidance for the teachers in primary school to know the most criteria that are appropriate for class conditions.

This paper begins with a literature review in Section 2, then followed by a review of the methodology used for conducting this study in Section 3. Section 4 presents the result of the study and the last part of the paper includes the conclusion and recommendation of the study.

## 2. LITERATURE REVIEW

### 2.1 STUDENT CHARACTERISTICS

Student characteristics is a personal quality of students who become characteristic and indicate the condition of students. This individual characteristic is believed to be a special ability that influences the degree of success in following a program [6]. Various information about student characteristics is needed by other learning components such as material goals, media, learning strategies and evaluation.

As describe first, the kind of student characteristics are intelligence levels, prior knowledge, cognitive styles, learning styles, motivation, and sosio-culture [1]–[3]. Each type of student characteristics has a different influence on determining the right instructional method as clearly describe on Figure 1.

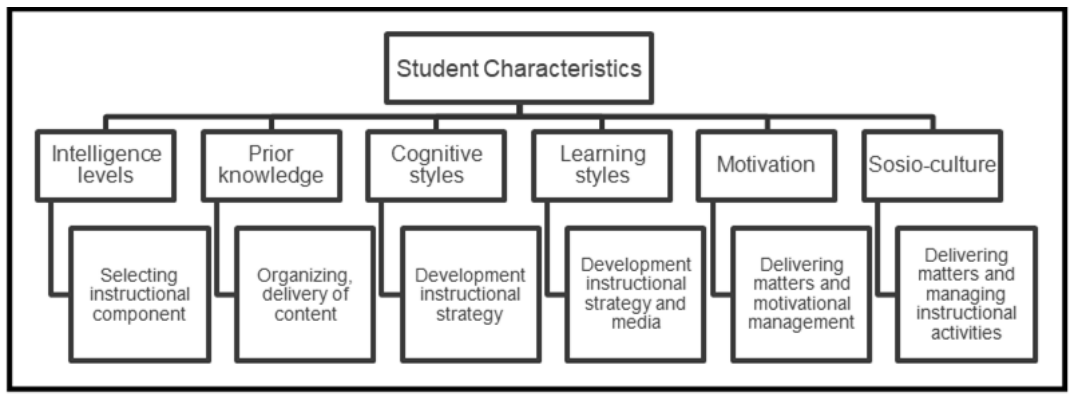

Figure 1. The influence of student characteristics on instructional method

Analyzing the characteristics of students can help in setting appropriate learning management strategies. Analysis of student characteristics can be done based on the dimensions of the characteristics of students who first adjusted to the objectives and characteristics of learning content. For example, the approach based on the basic characteristics of students (applying skills and abilities, experience in using technology), and learning context (personal identity, learning style, cognitive style, motivation) in formal and informal learning situations [2]. Student characteristics as a key interaction between students and learning that will affect the effectiveness of learning [7]. Therefore the teacher must recognize and consider the characteristics possessed by students proportionally so that learning can succeed well.

Intelligence levels can be described as the ability to solve problems, or to create products, that are valued within one or more cultural settings [8]. These are biopsychological potential. Whether and in what respects an individual may be deemed intelligent is a product of his genetic heritage and his psychological properties, ranging from his cognitive powers to his personality dispositions [1]. The level of intelligence of students can affect students' confidence in facing difficult material [7]. All students have various intelligence, or ways of receiving and expressing their knowledge. These intelligences can serve as a powerful springboard for creative curricular decision making and instructional planning [9].

The prior knowledge or mental models that learners bring to any information, therefore, can be vital to providing them with strategies and heuristics for managing the processing event [2]. It's important in increasing the meaningfulness of learning because it facilitates internal processes that take place within the students. Teachers often do not pay attention to how new knowledge is treated not related to prior knowledge. For example dividing facts and procedures as a separate material, emphasizing memorization and not understanding the material, and teaching managing new knowledge without understanding how the material reflects the student's personal goals or strategies in learning. Whereas linking with different types of abilities can make new knowledge meaningful to students to facilitate acquisition, organization, and re-disclosure of new knowledge that has been obtained. For the primary students, strengthening prior knowledge such as language, literacy, and self-regulation in various contexts proved to be most effective in increasing student success [10].

There are seven types of prior knowledge that could be used to facilitate the acquisition, organization, and re-disclosure of new knowledge [11]. Arbitrarily meaningful knowledge as a place to link memorized knowledge to facilitate retention. Analogic knowledge that associates new knowledge with other similar knowledge, but is beyond the content being discussed. Superordinate knowledge that can serve as a cantaloupe for new knowledge. Coordinate knowledge that can fulfill its function as associative and/or comparative knowledge. Subordinate knowledge that functions to concretize new knowledge or also provide examples. Experiential knowledge which also functions to concretize and provide examples for new knowledge. Cognitive strategy that provides ways of processing new knowledge, ranging from encoding, storage, to re-disclosure of knowledge that has been stored in memory.

Cognitive style refers to an individual's way of processing information [12] and reflect human attributes or processes that tend to be fixed over time [2]. It relates to things such as attention span, memory, mental procedures, and intellectual skills that determine how students feel, remember, think, solve problems, organize and represent information in their brain. Examples of cognitive styles include focussing and scanning, reflective and impulsive, serialist and wholist, and so on. Teachers must pay attention to the cognitive style of students in terms of choosing the right instructional strategy. The information of cognitive styles helping teachers to support media or learning experiences that match with the type of students cognitive styles so that student feel comfortable and equal with instructional activities on progress [5].

Learning style is defined as the characteristics, strengths, and tendencies that a person has towards how to receive and process information [13], and is related to the way a person begins to concentrate on processing and storing new and difficult information [14]. Learning styles are easier to alter over time with instruction, motivation, trial and error, and experience [2]. There are several models of learning style. Some of them are Kolb learning styles, Honey & Mumford models, V-A-R-K models, Myers-Briggs Type Indicator (MBTI) models, and Felder-Silverman models. Knowing student learning styles useful not only for the teachers in addition to design instructional with excellent quality but also for the students to more understanding their strength for learning optimally [5].

The same material can be accessed or understood in a different way by some people. So this means that it requires many ways so that the same information can be conveyed and understood by various people with different learning styles. Many of the methods referred to here are certainly related to the media in which the information is published and delivered through the right channels, as well as the choice of instructional strategies as a place to manage it.

As an example of material regarding a procedure that contains steps for operating a tool. Students with visual learning styles will more easily understand the material if it is presented through component drawings on the tool and the flow that must be done in its operation. Conversely, for students with verbal learning styles, it will be easier to understand the material in the form of oral descriptions or written texts that describe in detail the operating procedures of the tool. So that teachers must match learning strategies and learning tasks that are appropriate to the learning styles of students [15].

Student motivation on instructional is vital in ensuring that the learner persists adequately to successfully complete the task and acquire skills or content knowledge. Motivation to learn is identified by the student's choice of behavior, latency of the behavior, and the intensity and persistence of engagement in the learning task [16]. Motivation is not only limited to learning competence but also depends on the interaction of students with the social environment. This interaction is internalized in students from time to time and serves as a guide to behaving [1]. Therefore, identifying motivation should not only be related to the learning material but more to the overall attitudes and behavior of students towards their readiness to understand themselves and their environment in real terms.

Motivation has an impact on students' self-efficacy to succeed in learning. This gives a strong and positive influence on academic achievement [17]. When motivated enough, students tend to be more eager to face tasks, survive in difficult situations, and enjoy their achievements. There is a strong relationship between intrinsic motivation and student learning achievement. In addition, the learning context greatly influences student motivation. Organizing challenging learning materials, giving choices and learning autonomy have a positive impact on student motivation [16].

Socio-cultural conditions relate to groups or individuals associated with society. Examples of social characteristics are group structure, individual position in a society, social skills, and so on. Research indicates that the effect of socio-cultural conditions such as poverty, race/ethnicity, and community are distal factors associated with children's school success [10]. Students with good socio-cultural conditions can influence the success of learning in the long term [18]. A socio-cultural perspective must link theory with practice in instructional design that rests on the involvement of teachers and students in the real world environment with authentic practice communities [2]. Sociocultural instructional design views learning as collaborative and generative, involving enculturation into groups of students who are driven by authentic, ongoing, mediated, and built problems. Learning that takes into account sociocultural conditions emphasizes student-teacher collaboration so as to enable differences in individual learners, and the role of students' privileges in developing learning processes that are participatory, interactive and sustainable. Learning is not de-contextual, so tools are not seen as a delivery mechanism or as a tool that must be learned separately from the content of the material, but as a technology to add meaning to the process of learning and community development.

## 2.2 PRIMARY SCHOOL PROGRAMS

Primary school is the most basic level of formal education in Indonesia. Primary school is part of the basic education program launched by the government covering 6 years of primary school and 3 years of junior high school with a total of 9 years [19]. Primary schools are taken within 6 years, ranging from grades 1 to 6 with students ranging in age from 7 to 12 years.

The programs of primary school are focused on knowledge acquisition and learning process in general. Primary schools curriculum adheres to: (1) learning done by the teacher (taught curriculum) in the process developed in the form of learning activities in school, class, and community; and (2) students' direct learning experience (learned-curriculum) in accordance with the background, characteristics, and initial abilities of students. Individual direct learning experiences of students become learning outcomes for themselves while learning outcomes of all students become curriculum results [20].

Teachers who teach in primary school consist of class teachers and subject teachers. Class teachers teach language, math, natural science, social science, and cultural arts and crafts. The subject teachers teach religious education and physical education, sports and healthy subjects.

Primary school teachers must have the ability to working with people, organizing the classroom, planning the curriculum, managing behavior, assessing and record keeping, thinking about education, and becoming a professional [21]. At the same time must be able to manage the tension created by the need to support children's learning in the right way while responding to demands for accountability [22].

With the many demands that must be met by primary school teachers often cause stress and boredom that can affect the effectiveness of classroom learning. The teacher can no longer apply professionally, and just do the task as a teacher without feeling it. Therefore the teacher must be aware of his role which is vital to the student's success in learning.

## 2.3 RELATED WORKS

Recent research on student characteristics that influence learning outcomes has been widely discussed. Some of them are presented in Table 1.

Table 1. Types of characteristics of students discussed

| Author(s) | The main topics of student characteristics type | The result |
|---|---|---|
| Ergul, 2004 [17] | Motivation | Motivation has an impact on students' self-efficacy to succeed in learning. This gives a strong and positive influence on academic achievement. |
| Scheiter et.al, 2009 [7] | Prior Knowledge, Cognitive Styles, Intellectual Levels | Students who have a high level of prior knowledge tend to enjoy the learning process more. Students with a high level of intelligence are more confident in facing difficult material. Students who are able to adjust their cognitive style are able to demonstrate adaptive strategies using the information to overcome cognitive burden and demonstrate good problem-solving performance. |
| Kazu, 2009 [15] | Learning styles | Knowing learning styles will help teachers get to know student groups and provide more effective learning strategies. This has an impact on increasing student success and meaningful learning. |
| Kocakaya & Gonen,2012[18] | Demographic, Sosio-Cultural | Students with good social conditions can influence the success of learning in the long term. |
| McDonnald & Morrison, 2014 [10] | Prior Knowledge, Sosio-Cultural | Strengthening prior knowledge such as language, literacy, and self-regulation in various contexts proved to be most effective in increasing student success. The effect of socio-cultural conditions is distal factors associated with children's school success. |
| Surjono, 2015 [23] | Learning Styles | Students with certain learning style preferences will increase their learning achievement if the material is presented in a media that is in accordance with the characteristics of the learning style. |
| Rahadian & Budiningsih, 2017 [5] | Learning Styles | The teacher can accommodate students' learning styles by presenting strategies and learning media that are appropriate to their learning style preferences to get optimal learning outcomes. |

The research conducted focused only on some types of student characteristics, even though there were many types of student characteristics that could affect student achievement. The results of the study to provide good advice for teachers in addressing various types of characteristics of students they face. However, no one has shown the tendency of teachers to pay more attention to one type of characteristic of students who are considered the most important to be identified considering the various limitations they have, especially for primary school teachers. It would be helpful if there were studies that discussed the trends of types of student characteristics that were most considered by primary school teachers. So that need more study to find out the type of student characteristics that are most considered by primary school teachers and to know their effects on student learning achievement. In this sense, this work is new and significantly different from the previous study done in the field.

## 3. METHODOLOGY

### 3.1 CONTEXT AND PARTICIPANTS

Student characteristics is a personal quality of students who become characteristic and indicate the condition of students, which is believed to be a special ability that influences the degree of success in following a program [6]. Each type of student characteristics has a different influence on determining the right instructional method. Analyzing the characteristics of students can help in setting appropriate learning management strategies. Ideally, all types of student characteristics must be considered by the teachers because the variety of student characteristics influence the selection of other instructional method variables [5]. However various limitations causing them not to study in its entirety and depth. This study aims to find out the characteristics of students that most influence learning achievement so that it can help primary school teachers determine priorities according to class conditions. The population used in this study included 37 class teachers in grade 4 - 6 from seven primary schools.

### 3.2 PROCEDURE

This research was conducted from January 2018 – as the beginning of the semester – to March 2018 – as the middle of the semester – to get data about the planning carried out by the teacher and the learning outcomes that have been obtained by students. The teachers were given a questionnaire about the application of student characteristics in learning. Categorize the level of identification of student characteristics conducted by the teacher. Observe instructional planning documents from teachers that identify student characteristics. Categorize the level of application of information on the characteristics of students in the learning activities carried out by the teacher. Comparing students midterm exam to get a comparison of the value of learning outcomes between classes whose learning was developed based on students' characteristic information with classes that were not developed based on student characteristics information.

### 3.3 RESEARCH DESIGN

Completion of questionnaires on the application of student characteristics in learning must be ascertained the truth by observing instructional planning documents developed by the teacher. Categorizing the level of identification of characteristics of students by the teacher is intended to provide an overview of the types of student characteristics that are considered most important to be done by primary school teachers. Comparison of learning outcomes is intended to show improvement in learning outcomes that are developed based on information on student characteristics. The design of the study as shown in Figure 2.

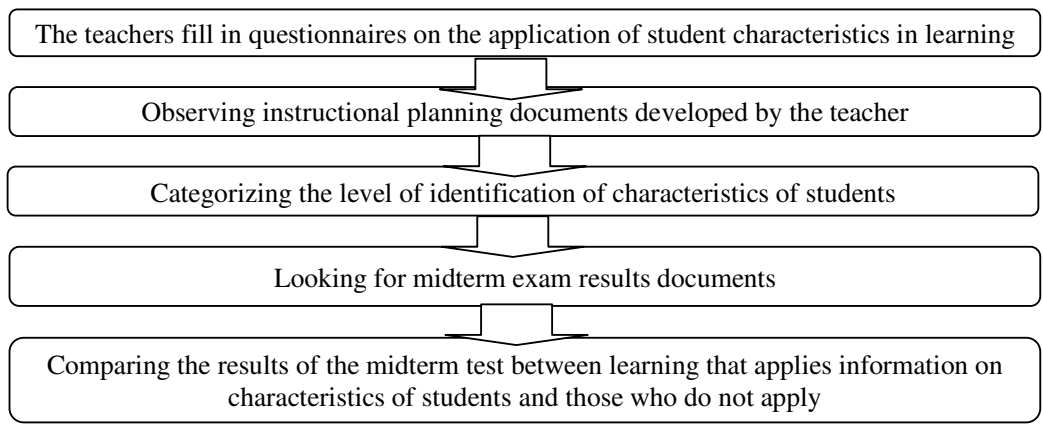

Figure 2. Design of study

A questionnaire which was adapted from [21], [24] used to find out the application of information on student characteristics in learning done by the teacher. The questionnaire, which was completed by each participant, consisted of four parts: (i) demographic information, (ii) knowledge of the types of students characteristics, (iii) how to identify student characteristics, (iv) development of appropriate strategies according to student characteristics, based on 5-point Likert scale (ranging from very disagree '1' to very agree '5') [25]. Data is presented in the tables form or frequency distributions. With this analysis will be known the tendency of the application of information characteristics of students at very low, low, medium, high, or very high levels. The conversion table of scale 5 can be seen in Table 2.

Table 2. Interval Conversion Average of Likert Scale

| Grade | Level | Score Interval | |
|---|---|---|---|
| A | Very High | $X > \overline{Xi} + 1,80\ SBi$ | $X > 4,21$ |
| B | High | $\overline{Xi} + 0,60\ SBi < X \leq \overline{Xi} + 1,80 SBi$ | $3,4 < X \leq 4,21$ |
| C | Medium | $\overline{Xi} - 0,60\ SBi < X \leq \overline{Xi} + 0,60\ SBi$ | $2,59 < X \leq 3,4$ |
| D | Low | $\overline{Xi} - 1,80\ SBi < X \leq \overline{Xi} - 0,60\ SBi$ | $1,79 < X \leq 2,59$ |
| E | Very Low | $X \leq \overline{Xi} - 1,80\ SBi$ | $X \leq 1,79$ |

Data on students learning achievement was obtained by questioning the teachers as well as from the average scores in the midterm exam. Test scores are separated into several categories as can be seen in Table 3.

Table 3. Category of Students Midterm Exam Scores

| Grade | Level | Score Interval |
|---|---|---|
| A | Very High | 86 - 100 |
| B | High | 71 – 85.99 |
| C | Medium | 56 – 70.99 |
| D | Low | 41 – 55.99 |
| E | Very Low | < 41 |

## 4. RESULTS

### 4.1 RESULT OF THE QUESTIONNAIRE

Before filling out the questionnaire, the teacher is given questions related to the types of characteristics of students identified in learning. The questions are as presented below:
Item number 1: What type of student characteristics that you identify in learning? Write down two or three types that you usually identify.

Item number 2: According to the answer of item number 1, choose one of the characteristics of students that you are most considering to identify.

The result of the item number 1 and 2 as shown in Table 4 and 5.

Table 4. Type of student characteristics that the teachers usually identify

| No | Type of Student Characteristic | Amount of teachers | Percentage |
|----|-------------------------------|--------------------|------------|
| 1 | Intelligence Level | 23 | 62% |
| 2 | Motivation | 17 | 46% |
| 3 | Cognitive Style | 12 | 32% |
| 4 | Learning Style | 12 | 32% |
| 5 | Prior Knowledge | 12 | 32% |
| 6 | Sosio-Cultural | 10 | 30% |

Table 5. Type of student characteristics that the teachers most considering to identify

| No | Type of Student Characteristic | Amount of teachers | Percentage |
|----|-------------------------------|--------------------|------------|
| 1 | Intelligence Level | 20 | 54% |
| 2 | Motivation | 6 | 16% |
| 3 | Cognitive Style | 4 | 11% |
| 4 | Learning Style | 4 | 11% |
| 5 | Prior Knowledge | 2 | 5% |
| 6 | Sosio-Cultural | 1 | 3% |

The results show that the types of student characteristics identified by primary teachers are the level of intelligence, motivation, cognitive style, learning style, prior knowledge, and socio-cultural. The type of student characteristics that are the most considered in learning is the level of intelligence.

Next, the teacher is given a questionnaire according to the type of characteristics that are most considered by them. The questionnaire used to get data about identifying the characteristics of students by the teachers. The number of respondents was 37 people divided into 20 teachers whose main consideration was the level of intelligence of students, 6 teachers on motivation, 4 teachers on cognitive style, 4 teachers in learning styles, 2 teachers in prior knowledge, and 1 teacher in socio-cultural. The results are described as follows:

Table 6. Questionnaire results by teachers who choose the student's intelligence level as the type most considered

| Statement | Strongly agree | Agree | Neutral | Disagree | Strongly disagree |
|---|---|---|---|---|---|
| I need to know the characteristics of all the students I teach | 45% | 55% | 0% | 0% | 0% |
| I am sure that the characteristics of students will influence their learning outcomes | 50% | 35% | 15% | 0% | 0% |
| I conducted a careful test in identifying the type of characteristics of students | 10% | 35% | 55% | 0% | 0% |
| I use a valid test tool to measure the type of characteristics of students | 10% | 35% | 55% | 0% | 0% |
| I use information on the type of student's characteristic in planning learning | 0% | 10% | 90% | 0% | 0% |
| I use the right learning component according to the type of student's characteristic | 0% | 10% | 70% | 20% | 0% |

Table 7. Questionnaire results by teachers who choose the student's motivation as the type most considered

| Statement | Strongly agree | Agree | Neutral | Disagree | Strongly disagree |
|---|---|---|---|---|---|
| I need to know the characteristics of all the students I teach | 0% | 50% | 50% | 0% | 0% |
| I am sure that the characteristics of students will influence their learning outcomes | 0% | 50% | 50% | 0% | 0% |
| I conducted a careful test in identifying the type of characteristics of students | 0% | 33% | 33% | 33% | 0% |
| I use a valid test tool to measure the type of characteristics of students | 0% | 33% | 33% | 33% | 0% |
| I use information on the type of student's characteristic in planning learning | 0% | 33% | 33% | 33% | 0% |
| I use the right learning component according to the type of student's characteristic | 0% | 33% | 50% | 17% | 0% |

Table 8. Questionnaire results by teachers who choose the student's cognitive style as the type most considered

| Statement | Strongly agree | Agree | Neutral | Disagree | Strongly disagree |
|---|---|---|---|---|---|
| I need to know the characteristics of all the students I teach | 25% | 25% | 25% | 25% | 0% |
| I am sure that the characteristics of students will influence their learning outcomes | 25% | 25% | 25% | 25% | 0% |
| I conducted a careful test in identifying the type of characteristics of students | 0% | 25% | 25% | 50% | 0% |
| I use a valid test tool to measure the type of characteristics of students | 0% | 25% | 25% | 50% | 0% |
| I use information on the type of student's characteristic in planning learning | 0% | 0% | 50% | 50% | 0% |
| I use the right learning component according to the type of student's characteristic | 0% | 0% | 50% | 50% | 0% |

Table 9. Questionnaire results by teachers who choose the student's learning styles as the type most considered

| Statement | Strongly agree | Agree | Neutral | Disagree | Strongly disagree |
|---|---|---|---|---|---|
| I need to know the characteristics of all the students I teach | 25% | 25% | 25% | 25% | 0% |
| I am sure that the characteristics of students will influence their learning outcomes | 0% | 50% | 25% | 25% | 0% |
| I conducted a careful test in identifying the type of characteristics of students | 0% | 25% | 25% | 50% | 0% |
| I use a valid test tool to measure the type of characteristics of students | 0% | 25% | 25% | 50% | 0% |
| I use information on the type of student's characteristic in planning learning | 0% | 0% | 50% | 50% | 0% |
| I use the right learning component according to the type of student's characteristic | 0% | 0% | 50% | 50% | 0% |

Table 10. Questionnaire results by teachers who choose the student's prior knowledge as the type most considered

| Statement | Strongly agree | Agree | Neutral | Disagree | Strongly disagree |
|---|---|---|---|---|---|
| I need to know the characteristics of all the students I teach | 0% | 50% | 50% | 0% | 0% |
| I am sure that the characteristics of students will influence their learning outcomes | 0% | 50% | 50% | 0% | 0% |
| I conducted a careful test in identifying the type of characteristics of students | 0% | 0% | 50% | 50% | 0% |
| I use a valid test tool to measure the type of characteristics of students | 0% | 0% | 50% | 50% | 0% |
| I use information on the type of student's characteristic in planning learning | 0% | 0% | 50% | 50% | 0% |
| I use the right learning component according to the type of student's characteristic | 0% | 0% | 50% | 50% | 0% |

Table 11. Questionnaire results by teachers who choose the socio-cultural level as the type most considered

| Statement | Strongly agree | Agree | Neutral | Disagree | Strongly disagree |
|---|---|---|---|---|---|
| I need to know the characteristics of all the students I teach | 0% | 0% | 100% | 0% | 0% |
| I am sure that the characteristics of students will influence their learning outcomes | 0% | 0% | 100% | 0% | 0% |
| I conducted a careful test in identifying the type of characteristics of students | 0% | 0% | 0% | 100% | 0% |
| I use a valid test tool to measure the type of characteristics of students | 0% | 0% | 0% | 100% | 0% |
| I use information on the type of student's characteristic in planning learning | 0% | 0% | 100% | 0% | 0% |
| I use the right learning component according to the type of student's characteristic | 0% | 0% | 0% | 100% | 0% |

## 4.2 CATEGORIZING THE LEVEL OF IDENTIFICATION STUDENT CHARACTERISTIC

To complete the data provided by the teacher, an observation was made of the instructional planning documents designed by the teacher. The results are converted using a Likert scale. Conversion of Likert scores in a scale of 5 is done to determine the level of identification of student characteristics by the teachers, as can be seen in Table 12.

Table 12. The level of identification of student characteristics carried out by the teachers

| No | Type of Student Characteristic | Score | Level |
|----|-------------------------------|-------|-------|
| 1 | Intelligence Level | 3.65 | High |
| 2 | Motivation | 3.19 | Medium |
| 3 | Cognitive Style | 2.92 | Medium |
| 4 | Prior Knowledge | 2.89 | Medium |
| 5 | Learning Style | 2.84 | Medium |
| 6 | Sosio-Cultural | 2.55 | Low |

As can be seen in Table 12, intelligence level occupies the highest level with a score of 3.65 which shows that primary school teachers are very considered about the student's intelligence level that they achieve in the implementation of learning. Furthermore, motivation ranks second, together with cognitive styles, prior knowledge, and learning styles at the medium level. This means that the teacher is enough to pay attention to the type of characteristics of these students in learning but not in depth. The last order is occupied by socio-cultural with a low level, which means that the teachers pay less attention to the socio-cultural conditions of students in learning.

### 4.3 RESULT OF THE MIDTERM EXAM

Although the level of intelligence ranks first in the type of student characteristics that are most important to be identified by the teacher in learning, the effect on student achievement is unknown. To find out whether there are differences in learning outcomes of students whose learning takes into account the level of intelligence of students and those who do not, it is necessary to compare the learning outcomes achieved by students. For this purpose, a document study is conducted to obtain the value data obtained by students on the midterm exam. The results of the comparison as presented in Table 13.

Table 13. Midterm Exam Score Comparison

| Instructional based on Students Intelectual Levels | Midterm Exam Score Per Subject | | | | | | Level |
|---|---|---|---|---|---|---|---|
| | Indonesian Language | Math | Social Science | Natural Science | Art-Culture-Craft- | Average | |
| No | 71.00 | 63.67 | 65.00 | 69.67 | 71.67 | **68.20** | Medium |
| Yes | 78.00 | 71.00 | 67.67 | 73.00 | 73.67 | **72.67** | High |
| Gain | 9% | 10% | 4% | 5% | 3% | 6% | |

Table 13 shows that the learning outcomes developed based on the identification of student's level of intelligence can be better. This can be seen from the acquisition of a better score on all subjects. Even in Indonesian language and mathematics subjects, the difference in learning outcomes is prominent.

## 5. DISCUSSION AND CONCLUSION

Student characteristic is believed to be a special ability that influences the degree of success in following a program [6]. Student characteristics as a key interaction between students and learning that will affect the effectiveness of learning [7]. Therefore the teacher must recognize and consider the characteristics possessed by students proportionally so that learning can succeed well. Ideally, all types of student characteristics must be considered by the teachers because the variety of student characteristics influence the selection of other instructional method variables [5]. However various limitations causing them not to study in its entirety and depth. This study aimed to find out what type of student characteristic most considered by teachers in primary school and to know their effects on student learning achievement. This is expected to help overcome primary school teacher confusion about which types of student characteristics are most considered in designing and implementing learning.

According to the result of the study, the type of student characteristics most considered by the teacher is student intelligence level. This can be seen from the majority of primary teachers who identify the types of characteristics of students. This can also be seen from the use of student characteristics information in instructional planning at a high level. Based on the study of learning planning documents developed by the teacher it was found that information on the level of intelligence of students was known from in-depth observation of students' behavior in solving learning problems and their speed in understanding the material presented.

The teachers use information on the student's intelligence levels to classify students into categories of fast, medium, and slow. Each category is treated differently so that students feel comfortable in understanding the material presented. Treatment differences can be in the form of material enrichment for fast category students and adding tasks to students in the fast category or organizing the right material for students in the slow category. Fast category students are given assignments as peers for their friends. Sometimes the explanation given by his friend is easier to understand. This will foster good learning motivation. Based on the comparison of learning outcomes it can be shown that students learning achievement which is developed based on the identification of students' intelligence level can be better than those who do not use it.

The future study should focus on the specific learning strategies are in accordance with the students intelligence level to help teachers choose the right strategy for their class conditions.

## ACKNOWLEDGEMENTS

The authors are thankful to Education and Cultural Offices of South Bangka, Islands of Bangka Belitung, Indonesia, for supporting this research work.

## AUTHORS

**Dwiwarna** is a senior primary school teacher in South Bangka Regency, Indonesia. Experienced as a school principal for several periods and now as a teacher position credit assessment team. He is interested in research on school management and instructional development.

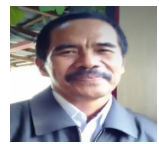

**Raditya Bayu Rahadian** hold his master's degree in instructional technology from Yogyakarta State University, Yogyakarta, Indonesia. He is an instructional technologist developer in South Bangka Regency, Indonesia since 2010. He is interested in research in the field of educational technology, including instructional development, educational services, and educational management.

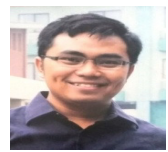